\documentclass[twocolumn,10pt]{article}

\usepackage[T1]{fontenc}
\usepackage[utf8]{inputenc}
\usepackage{graphicx}
\usepackage{amsmath}
\usepackage{amssymb}
\usepackage{siunitx}
\usepackage{threeparttable}
\usepackage{booktabs}
\usepackage{comment}
\usepackage{microtype}
\usepackage[colorlinks=true,allcolors=blue]{hyperref}
\usepackage[numbers,super,sort&compress]{natbib}
\usepackage[a4paper,top=2cm,bottom=2.5cm,left=1.5cm,right=1.5cm,columnsep=0.6cm]{geometry}
\usepackage{caption}
\usepackage{mathptmx}
\usepackage{fancyhdr}
\usepackage{titlesec}
\usepackage[T1]{fontenc}
\usepackage[utf8]{inputenc}  
\usepackage{xcolor}

\titleformat{\section}{\bfseries\large}{\thesection.}{0.5em}{}
\titleformat{\subsection}{\bfseries\normalsize}{\thesubsection.}{0.5em}{}
\titlespacing{\section}{0pt}{8pt}{4pt}
\titlespacing{\subsection}{0pt}{6pt}{2pt}

\begin{document}

\twocolumn[{%
\begin{center}

{\LARGE\bfseries
Emergent gold-induced interfacial magnetism in monolayer FePS$_3$}

\vspace{10pt}

{\normalsize
Giuseppe Buccoliero$^{1,2,*}$,
Anita Polli$^{1,3}$,
Marli dos Reis Cantarino$^{1}$,
Rachel Nickel$^{1}$,
Aloïs Arrighi$^{2}$,
Roberto Sant$^{4,1}$,
Tristan Riccardi$^{2}$,
Valérie Reita$^{2}$,
Nedjma Bendiab$^{2}$,
Laëtitia Marty$^{2}$,
Kurt Kummer$^{1}$,
Johann Coraux$^{2}$,
Nicholas B. Brookes$^{1}$\\
}
\vspace{6pt}
{\small
$^{1}$European Synchrotron Radiation Facility, 71 Avenue des Martyrs, CS 40220, 38043 Grenoble Cedex 9, France\\
$^{2}$Université Grenoble Alpes, CNRS, Grenoble INP, Institut NEEL, 38000 Grenoble, France\\
$^{3}$Department of Physics Giuseppe Occhialini, University of Milano-Bicocca, Piazza della Scienza 3, 20126 Milan, Italy\\
$^{4}$Université Grenoble Alpes, CEA, IRIG, MEM, 38000 Grenoble, France\\[4pt]
$^{*}$\texttt{giuseppe.buccoliero@esrf.fr}
}
\vspace{10pt}

\noindent\rule{\linewidth}{0.4pt}\\[4pt]
{\small\textbf{Abstract}\\[4pt]
\leftskip=0pt \rightskip=0pt \parfillskip=0pt plus 1fil\relax 
Owing to their atomically clean interfaces and extreme thinness, two-dimensional (2D) materials are ideal hosts for strong proximity effects, whereby contact with an adjacent material reshapes their ground state. Here, we demonstrate such an effect within a hybrid heterostructure, between the 2D antiferromagnet FePS$_3$ and an ultrathin gold film. Rather than a passive support, the latter profoundly changes the nature of the magnetism in the 2D magnet, inducing a sizeable interfacial moment. Large-area flakes from Au-assisted exfoliation, combined with element-selective X-ray absorption mapping, enable a layer-by-layer study from monolayer to bulk, benchmarked against an inert SiO$_2$ substrate reference. At 9~T and 5~K, the monolayer on gold shows a three-fold enhanced X-ray magnetic circular dichroism relative to SiO$_2$, with a $\sim$40\% reduction of the orbital-to-spin moment ratio. Resonant inelastic X-ray scattering reveals a collapse of the crystal-field ($d$--$d$) excitations and a massive spectral-weight transfer to the charge-transfer region, unveiling the microscopic mechanism behind this change. Supported by atomic multiplet calculations, we identify interfacial charge transfer and Fe~3$d$--Au hybridization as the origin of these effects, which decay over the first three layers and vanish at the pentalayer, establishing layer number combined with substrate engineering as an atomically precise handle.\\[4pt]}
\noindent\rule{\linewidth}{0.4pt}

\vspace{12pt}
\end{center}
}]

\section{Introduction}

\begin{figure*}[t]
    \centering
    \includegraphics[width=\textwidth]{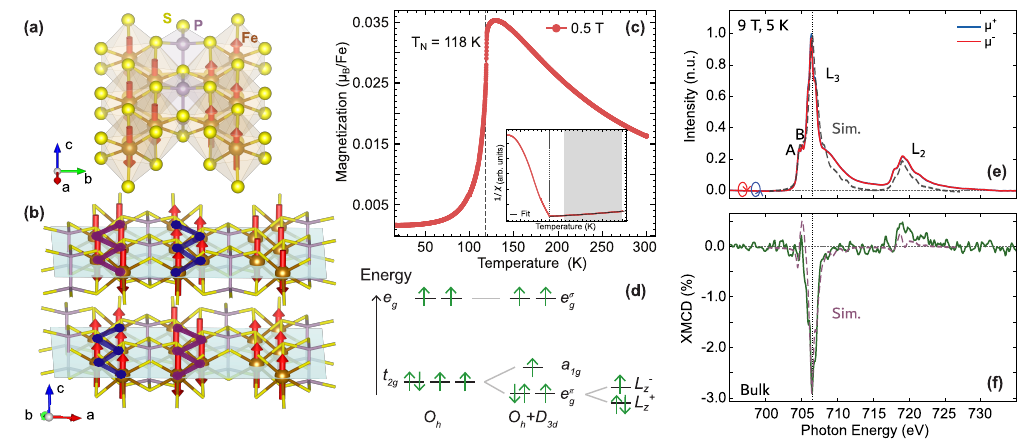}
    \caption{\textbf{Structure, bulk electronic and magnetic properties of FePS$_3$.} \textbf{a}~Side view (along the \textit{a} axis) of a single FePS$_3$ layer: Fe$^{2+}$ ions (brown) sit at the centre of edge-sharing FeS$_6$ octahedra bridged by P$_2$S$_6$ units, with S and P atoms shown in yellow and purple. Red arrows indicate the Fe$^{2+}$ magnetic moments, aligned along the easy axis perpendicular to the layers (\textit{c}). \textbf{b}~Side view of the bilayer illustrating the zigzag antiferromagnetic order: ferromagnetic chains carry moments along the easy axis, perpendicular to the layers ($c$, red arrows), with adjacent layers stacked antiferromagnetically (highlighted blue and purple chains oppositely aligned). \textbf{c}~Magnetization $M(T)$ of a bulk FePS$_3$ crystal measured under an out-of-plane magnetic field of 0.5~T, showing the antiferromagnetic onset at $T_\mathrm{N} \approx 118$~K (dashed line). Inset: inverse susceptibility $1/\chi$ with a Curie--Weiss fit (paramagnetic region shaded). \textbf{d}~Energy-level diagram of the Fe$^{2+}$ $3d^6$ manifold in the high-spin configuration under the octahedral crystal field ($O_h$), the trigonal distortion ($O_h+D_{3d}$), and spin--orbit coupling. \textbf{e,f}~Fe~$L_{3,2}$-edge XAS (e) and XMCD (f) of bulk FePS$_3$ at 9~T and 5~K in normal-incidence (NI) geometry, with beam and field along $c$. The two circular polarizations are shown as left ($\mu^{+}$) and right ($\mu^{-}$), with the atomic multiplet calculation overlaid in both panels (dashed line); peaks A and B mark the crystal-field pre-edge features.}
    \label{fig:fig1}
\end{figure*}

Magnetism in ultrathin films is remarkably sensitive to thickness, as established decades ago~\cite{Suter1979,Rau1989,Liu1990}. With traditional materials, however, exploring this sensitivity is challenging: the quality of the interfaces with the substrate, the capping layers and with it the ability to stabilize a well-defined thickness, are often critical issues~\cite{Vaz2008}. Van der Waals (vdW) magnets, in contrast, offer great stacking versatility, accommodating a broad range of substrates, while promising ultra-clean interfaces owing to the absence of dangling bonds at their surface. Accordingly, strongly thickness-dependent magnetism has been reported recently in a variety of compounds~\cite{Coraux2025,Park2026}, including a reduction of the ordering temperature~\cite{Gong2017,Gong2017_t}, changes in the nature of the magnetic order~\cite{huang2017layer}, and a dimensionality crossover in the statistical-physics sense~\cite{fei2018two}.

Within this landscape, the vdW semiconducting antiferromagnet iron phosphorus trisulfide (FePS$_3$), a member of the metal phosphorus sulfide family~\cite{samal2021two,kim}, stands out. It retains Ising-type antiferromagnetic (AFM) order down to the monolayer limit, with a N\'{e}el temperature that is almost independent of thickness, indicating that the weak interlayer coupling plays only a minor role in the magnetic ordering~\cite{lee2016ising,yang2024understanding,wang2016raman}. Indeed the material is characterized by critical exponents typical of a 2D Ising system, even in its bulk form~\cite{houmes2023magnetic}. Together with its Mott-like insulating character~\cite{tsurubayashi2018metal,jin2022mott}, unquenched orbital moments~\cite{dhakal2024hybrid,ovesen2026giant}, and giant single-ion anisotropy ($\sim$22~meV/Fe)~\cite{lee2023giant}, this makes FePS$_3$ an ideal platform to explore interfacial proximity effects. Its ground state is known to be influenced by the electronic environment: charge transfer from a WS$_2$ overlayer stabilizes an emergent ferromagnetic interfacial layer~\cite{gong2023ferromagnetism}, and non-magnetic molecular intercalation drives the bulk into a hard-ferromagnetic state~\cite{ou2025hard}.

Element-specific synchrotron spectroscopies are uniquely suited to understand magnetic properties in the light of the details of the local microscopic electronic structure. Specifically, X-ray magnetic circular dichroism (XMCD)~\cite{xmcd,thole1992x,carra1993x} and resonant inelastic X-ray scattering (RIXS)~\cite{ament2011rmp,de2008core} have provided fundamental insight into related vdW magnets~\cite{mitrano2024exploring,buccoliero2026ligand}, and start to be increasingly applied to exfoliated few-layer samples~\cite{fujita2022x} to separate dimensionality from bulk electronic structure, as in NiPS$_3$~\cite{discala2024elucidating}. In bulk FePS$_3$, RIXS has resolved the spin-orbital multiplet excitations that encode the magnetic phase transition, persisting in exfoliated flakes down to five layers~\cite{wei2025spin}. Reaching the genuine 2D limit, however, remains challenging: flakes, produced by mechanical exfoliation, have limited lateral extension~\cite{desai2016gold}---often smaller than the footprint of soft X-ray synchrotron beams---and relatively poor surface cleanliness. Additionally, finding small flakes across the macroscopic surface of a substrate can be practically challenging in the experimental setups where these techniques are implemented.

Here we overcome these constraints by combining Au-assisted mechanical exfoliation --- which exploits the strong Au--S chemical affinity to produce large-area monolayer flakes with lateral dimensions of hundreds of micrometres~\cite{huang2020universal,velicky2018mechanism,desai2016gold} --- with X-ray absorption mapping to locate and identify them, enabling a systematic, layer-by-layer XMCD study of FePS$_3$ from monolayer to bulk on Au(2--4~nm)/Ti(2~nm)/SiO$_2$ substrate and, as an inert reference, on a bilayer exfoliated onto bare SiO$_2$. This gives access to the element-specific spin and orbital moments of FePS$_3$ across the whole thickness range, and pushes RIXS on this compound from the five-layer limit reached so far~\cite{wei2025spin} down to the single layer. We find that gold is not a passive support: rather, it drives a charge-transfer reconstruction of the Fe$^{2+}$ electronic structure and magnetic ground state, yielding a remarkably enhanced XMCD response. Absent on the inert SiO$_2$ reference substrate, the signal decays over the first three layers, with bulk AFM behaviour restored by five. Atomic multiplet calculations (AMCs)~\cite{quanty,haverkort} reproduce the localized-limit spectra of the bulk and of the flake on SiO$_2$, providing the reference against which the gold-induced response stands out as a departure from the localized multiplet regime. RIXS validates this picture directly: a suppression of the crystal-field excitations and a pronounced enhancement of the charge-transfer excitations, observed exclusively on gold, mark a reconstruction of the Fe$^{2+}$ local electronic structure that underpins this emergent, enhanced magnetic state.

\section{Results}

\subsection{Electronic and magnetic structure of bulk FePS$_3$}

\begin{figure*}[h]
    \centering
    \includegraphics[width=\textwidth]{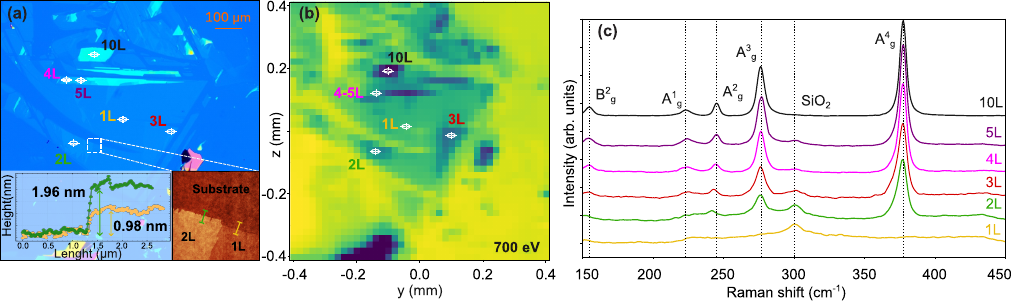}
    \caption{\textbf{Thickness characterization and identification of FePS$_3$ flakes.} \textbf{a}~Optical micrograph of S3; white targets mark the spots where Raman spectra were acquired, labelled by thickness (1L to 10L; scale bar 100~$\mu$m). Right inset: AFM image of the 1L/2L edge. Left inset: AFM height profile across the 1L/2L boundary, as indicated on the AFM image, showing single- and double-layer steps. \textbf{b}~Off-resonance (700~eV) X-ray absorption map of the same region, with the same spots labelled. \textbf{c}~Raman spectra measured at the marked spots (300~K), from 1L to 10L.}
    \label{fig:fig2}
\end{figure*}

The robust Ising-like antiferromagnetic order of FePS$_3$ emerges from Fe$^{2+}$ ions arranged on a honeycomb lattice within each vdW layer, at the centre of edge-sharing FeS$_6$ octahedra bridged by P$_2$S$_6$ units [Fig.~\ref{fig:fig1}(a)]~\cite{lanccon2016magnetic,xie2019crystallographic}. Below the N\'{e}el temperature the spins order in a zigzag antiferromagnetic pattern: ferromagnetic chains along one crystallographic direction couple antiferromagnetically to their neighbours, with moments along the magnetic easy axis, perpendicular to the layers ($c$), as shown for the bilayer in Fig.~\ref{fig:fig1}(b), where adjacent layers stack antiferromagnetically. The bulk transition is apparent in the magnetization $M(T)$ curve which displays the characteristic cusp at $T_\mathrm{N} \approx 118$~K [Fig.~\ref{fig:fig1}(c)], in agreement with previous reports~\cite{lee2016ising}; a Curie--Weiss fit to the inverse susceptibility [inset in Fig.~\ref{fig:fig1}(c)] yields a paramagnetic temperature $\theta \approx -81$~K, confirming the dominant antiferromagnetic interactions~\cite{joy1992magnetism}. 

These magnetic properties are rooted in the local electronic structure of the Fe$^{2+}$ ion ($3d^6$ configuration)~\cite{piacentini1982x} in its octahedral ($O_h$) environment, lowered by a trigonal ($D_{3d}$) distortion arising from the elongation of the FeS$_6$ octahedron along its threefold axis~\cite{joy1992magnetism,lanccon2016magnetic}. The octahedral crystal field splits the $3d$ states into a lower $t_{2g}$ and an upper $e_g^{\sigma}$ manifold; the trigonal distortion further splits the $t_{2g}$ states into a lower $e_g^{\pi}$ doublet and an upper $a_{1g}$ singlet~\cite{wei2025spin}. Spin--orbit coupling then lifts the remaining orbital degeneracy of the low-lying states, splitting them into $L_z^{+}$ and $L_z^{-}$ components~\cite{yang2024understanding} --- the microscopic origin of the unquenched orbital moment and giant single-ion anisotropy of FePS$_3$~\cite{lee2023giant}[Fig.~\ref{fig:fig1}(d)].

This $3d$ manifold is directly reflected in the X-ray absorption [Fig.~\ref{fig:fig1}(e)] and dichroism [Fig.~\ref{fig:fig1}(f)] of bulk flakes. XMCD measures the net magnetic moment projected along the beam and thus vanishes for a perfectly compensated collinear antiferromagnet: it probes the residual sublattice imbalance, which the applied field can only weakly perturb here, since bulk FePS$_3$ magnetizes along its easy axis only above $\sim$35~T~\cite{wildes2020high}. The Fe~$L_{3,2}$-edge XAS exhibits a pronounced multiplet structure with a characteristic double pre-edge feature (peaks A and B), corresponding to $2p \rightarrow 3d$ transitions into distinct multiplet final states whose separation is set by the trigonal splitting. At 9~T and 5~K in normal-incidence (NI) geometry, the bulk shows only a weak XMCD signal of $\sim$3\% [Fig.~\ref{fig:fig1}(f)], consistent with the antiferromagnetic ground state, the dichroism arising from a small field-induced uncompensation of the sublattice moments. Both lineshapes are well reproduced by AMCs using parameters consistent with related work~\cite{wei2025spin}; the multiplet structure is particularly sensitive to the hybridization potentials and to the charge-transfer energy. The calculated ground state carries an unquenched orbital moment $m_l \approx 1.05~\mu_B$ and a spin moment $m_s \approx 3.56~\mu_B$ per Fe atom ($m_l/m_s \approx 0.29$), as expected for Fe$^{2+}$ in this trigonally distorted environment~\cite{ovesen2026giant,dhakal2024hybrid}. A sum-rule analysis~\cite{thole1992x,carra1993x} of the measured XMCD gives $m_l/m_s \approx 0.33$, in agreement with the AMCs value and used as the bulk reference throughout.

Consistent with this large anisotropy, the antiferromagnetic order persists up to 9~T, with no sign of a metamagnetic transition (spin-flop or spin-flip) in either the XMCD response or in bulk $M(H)$ curves measured by magnetometry~\cite{wildes2020high}, in contrast to related MPS$_3$ compounds with quenched orbital moments and lower anisotropy, where such transitions occur at lower fields (e.g.\ CrPS$_4$~\cite{adv,buccoliero2026ligand}, MnPS$_3$~\cite{basnet2021highly,MnPS3}).

\subsection{Gold-induced proximity effects}

Thickness-dependent measurements were performed on flakes exfoliated onto a gold-coated substrate. Our flakes had a lateral extension of the order of hundreds of micrometres, and an ultra-clean surface essentially free of adhesive-tape residue (see Methods)~\cite{desai2016gold,huang2020universal,velicky2018mechanism}. Sample thickness was  characterized by atomic force microscopy (AFM) and Raman scattering spectroscopy, the latter providing thickness-specific spectral fingerprints comparable across samples~\cite{lee2016ising,wang2016raman}. Further details on flake localization, air stability, and beam damage are reported in the Supplementary Information~\cite{ramos2021ultra,jimenez2024x}. A bilayer flake ($\sim$70~$\mu$m$^2$), exfoliated without gold onto bare SiO$_2$, serves as a thickness-matched reference to disentangle the intrinsic few-layer response from the proximity effect of the metallic film. The three measured samples are summarized in Table~\ref{tab:samples}.

\begin{table}[htbp]
    \centering
    \caption{Summary of the measured samples.}
    \label{tab:samples}
    \footnotesize
    \setlength{\tabcolsep}{4pt}
    \renewcommand{\arraystretch}{1.25}
    \begin{tabular}{@{}cllc@{}}
        \toprule
        Sample & Substrate & Exfoliation & Layers \\
        \midrule
        S1 & SiO$_2$/Si                  & Glovebox & 2    \\
        S2 & Au(4\,nm)/Ti(2\,nm)/SiO$_2$ & Glovebox & 1 \\
        S3 & Au(2\,nm)/Ti(2\,nm)/SiO$_2$ & Air      & 1--5 \\
        \bottomrule
    \end{tabular}
\end{table}

An optical-microscope image of sample S3 and an off-resonance (700~eV) X-ray absorption map of the same region are shown in Fig.~\ref{fig:fig2}(a--b). The correspondence between X-ray absorption maps and optical images allows the beam (FWHM $\sim$10$\times$20~$\mu$m$^2$) to be positioned on spots of different, well-defined thickness, as indicated in Fig.~\ref{fig:fig2}(b) for S3. AFM profiles identifying the monolayer and bilayer regions are shown as an inset in Fig.~\ref{fig:fig2}(a): relative to the substrate roughness ($\sim$0.2~nm), the measured step heights are below 1 and 2~nm, respectively, consistent with the reported interlayer spacing of FePS$_3$~\cite{du2016weak,lee2016ising}. Combining AFM with Raman scattering spectroscopy on the same spots provides an unambiguous thickness assignment: as illustrated in the Raman cascade of Fig.~\ref{fig:fig2}(c), the FePS$_3$ phonon modes between 150 and 450~cm$^{-1}$ --- labelled $B_g^2$, $A_g^1$, $A_g^2$, $A_g^3$, and $A_g^4$~\cite{yan2022layer} and arising from the vibrations of the [P$_2$S$_6$]$^{4-}$ units and the Fe sublattice --- weaken and shift with decreasing thickness, fully consistent with the literature~\cite{yan2022layer,wang2016raman,du2016weak}. The most intense mode, $A_g^4$ near 380~cm$^{-1}$, serves as the main thickness marker, while the broad feature near 300~cm$^{-1}$ originates from the SiO$_2$/Si substrate and is therefore most prominent for the thinnest flakes. In particular, the $A_g^2$ mode red-shifts from 245~cm$^{-1}$ in bulk-like flakes to 242~cm$^{-1}$ in the monolayer, as similarly reported~\cite{lee2016ising}.

\begin{figure}[t]
    \centering
    \includegraphics[width=\columnwidth]{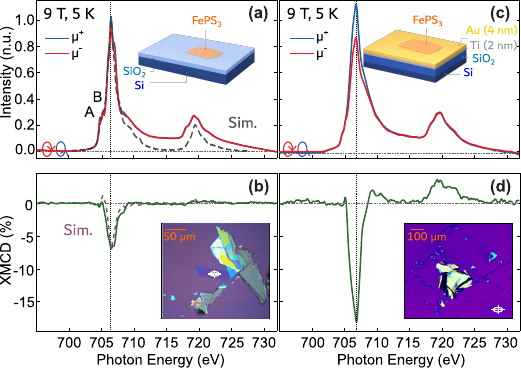}
    \caption{\textbf{FePS$_3$ on inert SiO$_2$ versus on gold.} \textbf{a,b}~Fe~$L_{3,2}$-edge XAS (a) and XMCD (b) of the bilayer on SiO$_2$ (S1) at 9~T and 5~K in NI geometry. The two circular polarizations are shown as left ($\mu^{+}$) and right ($\mu^{-}$), with the atomic multiplet calculation overlaid in both panels (dashed line); peaks A and B mark the crystal-field pre-edge features. The insets show a schematic of the bilayer on SiO$_2$/Si (a) and an optical micrograph of the flake, where the white target marks the beam position and footprint (b, scale bar 50~$\mu$m). \textbf{c,d}~Corresponding XAS (c) and XMCD (d) of the monolayer on Au(4~nm)/Ti(2~nm)/SiO$_2$ (S2) at 9~T and 5~K in NI geometry. The insets show a schematic of the monolayer on gold (c) and an optical micrograph of the flake, where the white target marks the beam position and footprint (d, scale bar 100~$\mu$m).}
    \label{fig:fig3}
\end{figure}

Figures~\ref{fig:fig3}(a--b) show the XAS and XMCD spectra of the FePS$_3$ bilayer on SiO$_2$ (S1) (9~T, 5~K, NI geometry). The XAS lineshape resembles that of the bulk --- the pre-edge multiplet (peaks A and B) is retained, though weaker and broader --- with no measurable chemical shift of the $L_3$ edge, indicating that the Fe$^{2+}$ valence is preserved. The XMCD signal, in contrast, is roughly doubled relative to the bulk ($\sim$6\%), indicating a net out-of-plane moment on the Fe$^{2+}$ sites, consistent with the field-induced uncompensated magnetization that spin Hall magnetoresistance has revealed at FePS$_3$ surfaces for both in-plane and out-of-plane applied fields~\cite{feringa2022observation}. The corresponding $m_l/m_s$ ratio is enhanced relative to the bulk ($\approx$0.37), pointing to a still-localized Fe$^{2+}$ multiplet that retains the large orbital contribution of the atomic bulk ground state~\cite{ovesen2026giant,lee2023giant}. The same trend was reported for exfoliated CrSiTe$_3$ approaching the monolayer~\cite{fujita2022x}, where it was attributed to reduced dimensionality and, possibly, to the magnetic dipole term $\langle T_z\rangle$, which our sum-rule analysis neglects and which grows once the local symmetry is lowered~\cite{crocombette1996importance}. Our AMCs support this, giving a $\langle T_z\rangle$ that is opposite in sign to $\langle S_z\rangle$ and grows by $\sim$20\% from bulk to bilayer (see Supplementary Information).  

Two complementary refinements of the atomic multiplet model capture this response. The enhanced XMCD amplitude is reproduced by unbalancing the two antiferromagnetic sublattices (see Methods), while the XAS lineshape is matched by a moderate reduction of the charge-transfer energy and an anisotropic redistribution of the metal--ligand hybridization --- both fingerprints of a truly 2D behaviour. Crucially, both effects remain within the localized multiplet regime: the AMC still holds on inert SiO$_2$, identifying this response as a genuine dimensionality effect and setting the reference against which the gold-induced reconstruction will stand out.

On the Au substrate (S2) the XAS lineshape measured under identical conditions changes radically: the multiplet pre-edge peaks are entirely lost and the spectrum acquires a markedly broader character [Fig.~\ref{fig:fig3}(c)].
This is accompanied by a strongly enhanced XMCD signal, as much as 18.5\% [Fig.~\ref{fig:fig3}(d)], three times larger than on the SiO$_2$ reference. Notably, the $L_3$ edge does not shift significantly in energy, showing that the reconstruction is electronic in nature and does not proceed through Fe oxidation. The $m_l/m_s$ ratio is instead reduced ($\approx$0.19) relative to the bulk, signalling a partial quenching of the orbital moment. The opposite evolution of $m_l/m_s$ on the two substrates --- enhanced on SiO$_2$, reduced on Au --- points to two distinct mechanisms: a field-induced uncompensation of localized moments in the former, against an interface-driven charge transfer in the latter. Consistently, the broad metallic-like XAS lineshape of S2 falls outside the reach of the AMCs, which in essence describe transitions between localized multiplet states.

\begin{figure*}[!htbp]
    \centering
    \includegraphics[width=\textwidth]{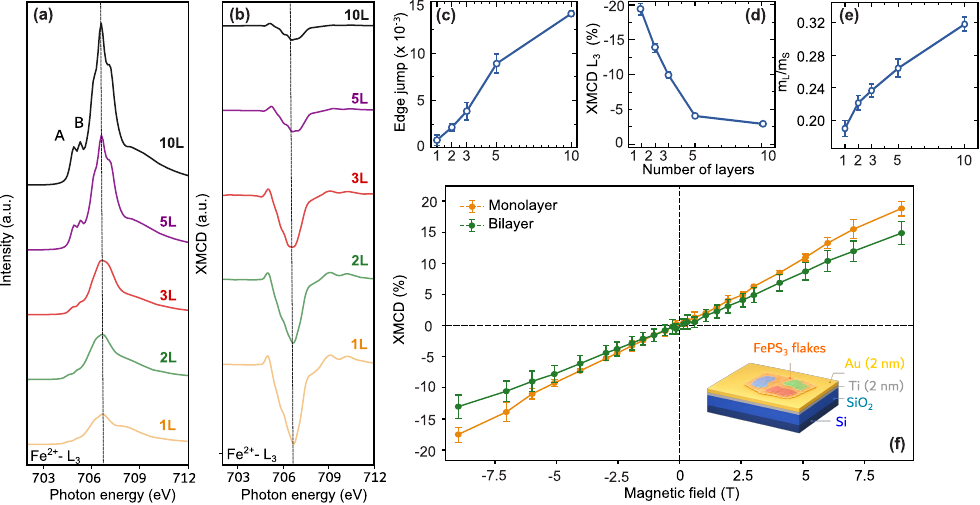}
    \caption{\textbf{Layer-dependent decay of the interfacial magnetism in FePS$_3$ on gold (sample S3).} \textbf{a,b}~Fe~$L_3$-edge XAS (a) and XMCD (b) cascades from the monolayer (1L) to 10L, measured at 9~T and 5~K in normal-incidence geometry; the dashed line marks the Fe$^{2+}$ $L_3$ resonance, and peaks A and B mark the pre-edge multiplet features, progressively recovered with increasing thickness. \textbf{c--e}~Thickness dependence of the XAS edge jump (c), of the XMCD amplitude at the Fe$^{2+}$ $L_3$ peak (d), and of the $m_l/m_s$ ratio extracted from sum rules (e). \textbf{f}~Element-specific magnetization curves of the monolayer (orange) and bilayer (green), measured at the Fe$^{2+}$ $L_3$ peak energy while sweeping the field from $-9$ to $+9$~T at 5~K. Error bars are estimated from the residual field-independent contribution to the dichroism (see Methods). Inset: schematic of the FePS$_3$ flakes on Au(2~nm)/Ti(2~nm)/SiO$_2$/Si.}
    \label{fig:fig4}
\end{figure*}

\begin{figure}[t]
    \centering
    \includegraphics[width=\columnwidth]{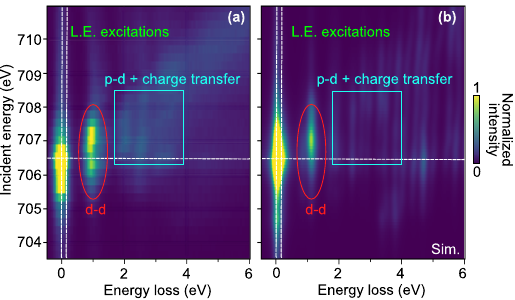}
    \caption{\textbf{RIXS energy map of bulk FePS$_3$ at the Fe~$L_3$ edge.} \textbf{a}~Experimental RIXS intensity as a function of incident energy and energy loss, normalized to its maximum, acquired at 5~K and fixed transferred momentum ($|\mathbf{q}|\approx0.51$~\AA$^{-1}$) with the average of horizontal- and vertical-linearly polarized light. The vertical dashed white lines mark the elastic line and the low-energy (L.E.) excitations near zero loss; the $d$--$d$ excitations around 1~eV (red ellipse) and the LMCT ($pd$) and charge-transfer region at 2--4~eV (cyan box) are also highlighted. The horizontal dashed line marks the resonant $L_3$ incident energy. \textbf{b}~Corresponding atomic multiplet calculation obtained with the same bulk parameter set.}
    \label{fig:fig5}
\end{figure}

\subsection{Layer-dependent decay of the interfacial magnetism}

The evolution of the magnetic signal with thickness was then probed on S3, from one to five layers and up to tens of layers (bulk-like). Figure~\ref{fig:fig4}(a) shows a XAS cascade revealing a progressive evolution of the lineshape, from the monolayer --- metallic-like, without a resolved pre-edge --- toward the five-layer flake, where the pre-edge peaks A and B are essentially recovered; a weak pre-edge multiplet feature already begins to reappear from two to three layers. Figure~\ref{fig:fig4}(b) shows the corresponding XMCD response, which decreases progressively from the monolayer to five layers. Importantly, the enhancement is not confined to the contact layer: the bilayer still exhibits a sizeable XMCD of $\sim$14\%, more than twice that of the bilayer/SiO$_2$ (S1) under identical conditions, and only at five layers and above does the signal return to the bulk value. This gradual, rather than abrupt, decay indicates that the interfacial reconstruction, while strongest at the Au-contacted layer, extends over the first few layers before bulk antiferromagnetic compensation is restored.

Sample S3 was exfoliated in air and immediately transferred to the glovebox. The XMCD magnitude and lineshape are unaffected by this exposure; in some monolayer regions, however, an additional post-edge XAS feature appears --- more pronounced than for thicker flakes --- which we attribute to a minor amount of iron oxide, detectable owing to the extreme element sensitivity of the technique. This feature does not affect the magnetic response: it lies at an energy distinct from the Fe$^{2+}$ peak used for the XMCD, is spatially heterogeneous rather than systematic, and contributes negligibly to the dichroism at 9~T. Crucially, the air-exfoliated series (S3) and the glovebox-exfoliated sample (S2) --- with 2 and 4~nm of gold, respectively --- yield XMCD signals of consistent magnitude, with coinciding monolayer $m_l/m_s$ ratios (0.19). This agreement shows that neither the brief air exposure nor the gold thickness affects the reconstruction, in line with our Raman scattering data (see Supplementary Information) and with previous reports on the air stability of FePS$_3$~\cite{ramos2021ultra,jimenez2024x}.

The overall trends are summarized in Fig.~\ref{fig:fig4}(c--e): the XAS edge jump recovers with increasing thickness, being about 15 times smaller in the monolayer than in the bulk [Fig.~\ref{fig:fig4}(c)]; the XMCD amplitude decreases toward the bulk value [Fig.~\ref{fig:fig4}(d)]; and the $m_l/m_s$ ratio rises from its reduced monolayer value and saturates toward the bulk value at higher thickness [Fig.~\ref{fig:fig4}(e)]. This rise tracks the evolution of the XAS and XMCD lineshapes: as the first layers acquire a more itinerant character, the orbital moment is progressively quenched. Finally, Fig.~\ref{fig:fig4}(f) shows the element-specific magnetization curves measured on S3 for the monolayer and the bilayer at the Fe$^{2+}$ $L_3$ peak energy ($\sim$706.7~eV). Both curves are linear and hysteresis-free up to 9~T, ruling out spontaneous ferromagnetic order and pointing instead to a field-induced, uncompensated magnetization --- a field-proportional, susceptibility-like response rather than saturating local moments. Notably, the bilayer response is only $\sim$25\% smaller than that of the monolayer, well short of the $\sim$50\% reduction expected if the reconstruction were confined to the single Au-contacted layer, indicating that the second layer still carries a substantial magnetic moment. This susceptibility-like response, strongly amplified at the Au interface and decaying back to the bulk antiferromagnetic state within the first three layers, sets the stage for the mechanism discussed below.

\subsection{Interfacial charge transfer probed by RIXS}
RIXS gives direct access to the effect of both reduced dimensionality and the metallic substrate on the local electronic structure of FePS$_3$~\cite{ament2011rmp,de2008core}. Being a photon-in--photon-out, element-specific spectroscopy, it probes the charge-neutral $d$--$d$ and charge-transfer excitations of the Fe$^{2+}$ multiplet~\cite{wei2025spin}, making it ideally suited to track the interfacial reconstruction.

A RIXS energy map around the Fe$^{2+}$~$L_{3}$ edge is shown in Fig.~\ref{fig:fig5}(a) for bulk FePS$_3$, resolving both the crystal-field and the ligand-to-metal charge-transfer (LMCT) excitations, in agreement with previous reports~\cite{wei2025spin}: split $d$--$d$ excitations around 1~eV energy loss, followed by the LMCT ($pd$) excitations and a broader charge-transfer region. The map is reproduced by the AMCs [Fig.~\ref{fig:fig5}(b)] with the same parameter set used for the XAS and XMCD, demonstrating a consistent description across different spectroscopies. Below the $d$--$d$ manifold, two low-energy excitations are expected in addition to the elastic line: a spin--orbit exciton around 100~meV, not resolved at the present energy resolution, and a peak around 220~meV arising from the trigonal splitting of the ${}^{5}T_{2g}$ ground state (the $e_g^{\pi}\rightarrow a_{1g}$ transition), whose energy reflects the magnitude of the trigonal distortion. The energy gap between this distortion peak and the first $d$--$d$ excitation is governed by the on-site Coulomb and exchange interactions ($U_{dd}\approx4.5$~eV), the metal--ligand covalency, and the crystal-field splitting, reflecting the correlated, charge-transfer/Mott--Hubbard insulating character of the bulk~\cite{zaanen1985band}.

\begin{figure}[t]
    \centering
    \includegraphics[width=\columnwidth]{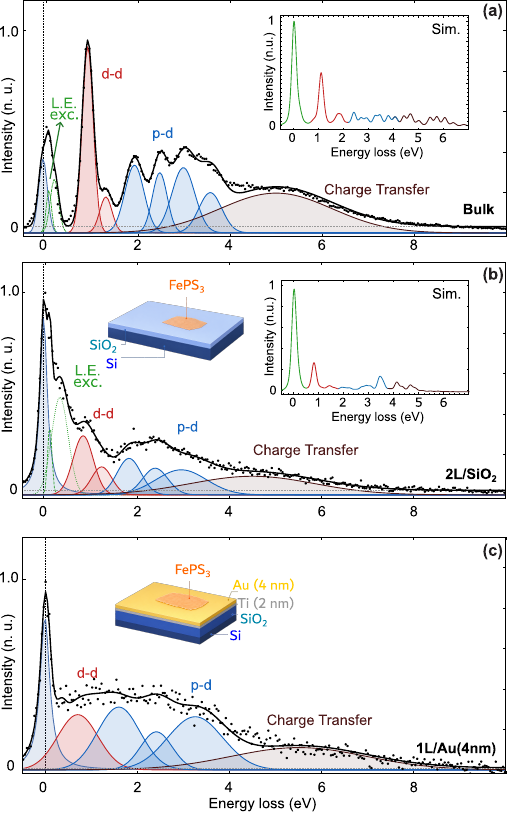}
    \caption{\textbf{RIXS spectra at the Fe~$L_3$ edge for bulk and few-layer FePS$_3$.} \textbf{a}~Bulk, \textbf{b}~bilayer on SiO$_2$/Si (S1), and \textbf{c}~monolayer on Au(4~nm)/Ti(2~nm)/SiO$_2$ (S2), all measured at 5~K with horizontal incident polarization, at an incident energy of 706.7~eV and fixed transferred momentum ($|\mathbf{q}|\approx0.51$~\AA$^{-1}$). Experimental data (black dots) are shown with the total fit (black line) and its spectral decomposition into the low-energy (L.E.) excitations (green), the $d$--$d$ excitations (red), the LMCT ($pd$) excitations (blue), and the charge-transfer continuum (brown). Insets in \textbf{a} and \textbf{b}: corresponding atomic multiplet calculations, obtained with the parameter sets of Table~\ref{tab:quanty}. Insets in \textbf{b} and \textbf{c}: schematics of the measured samples.}
    \label{fig:fig6}
\end{figure}

To resolve the electronic features more clearly, RIXS spectra were measured at the Fe$^{2+}$~$L_3$ edge, which also provide a better signal-to-noise ratio for the thinnest flakes, where the low counting rate is the main limiting factor. Figure~\ref{fig:fig6}(a) shows a fit of a bulk spectrum with the corresponding AMC (inset): in order of increasing energy loss, the spin--orbit exciton and the trigonal-distortion peak introduced above, followed by the main ${}^{5}T_{2g}\rightarrow{}^{5}E_{g}$ $d$--$d$ excitation, split into two components at $\sim$1 and $\sim$1.3~eV by the dynamical Jahn--Teller coupling of the orbitally degenerate ${}^{5}E_{g}$ excited state~\cite{wei2025spin,huang2017jahn}. At higher energy loss, peaks centred between 1.8 and 3.6~eV are attributed to LMCT ($pd$) excitations, followed, beyond $\sim$4~eV, by a broad, fluorescence-like charge-transfer continuum. All these features persist unchanged in exfoliated flakes down to five layers~\cite{wei2025spin}, consistent with our XAS and XMCD, which recover bulk-like electronic and magnetic behaviour at that thickness.

This picture changes significantly for the bilayer measured on the bare-SiO$_2$ sample (S1), whose spectrum is reported in Fig.~\ref{fig:fig6}(b). The same excitations remain resolved --- the spin--orbit exciton, the trigonal-distortion peak, and the ${}^{5}T_{2g}\rightarrow{}^{5}E_{g}$ manifold --- but they are systematically broader and weaker, while the LMCT ($pd$) features broaden and gain spectral weight, becoming as intense as the crystal-field excitations, and the charge-transfer region broadens with a reduced gap, signalling an increased covalent character. This behaviour closely mirrors that of exfoliated NiPS$_3$ on bare SiO$_2$~\cite{discala2024elucidating}, where reduced dimensionality drives a predominantly electronic modification of the RIXS spectra: the charge-transfer energy $\Delta$ decreases faster than the metal--ligand hopping $t$, so that the effective, ligand-mediated width of the $3d$ manifold ($\propto t^2/\Delta$~\cite{zaanen1985band}) increases, shifting the ground state toward a more covalent character and enhancing the charge-transfer spectral weight. Our multiplet model captures the same trend, reproducing both the XAS [Fig.~\ref{fig:fig3}(a)] and the RIXS spectra of the SiO$_2$ bilayer [Fig.~\ref{fig:fig6}(b), inset] through a reduction of the charge-transfer energy (from 1.6 to 0.8~eV) accompanied by an anisotropic redistribution of the hybridization potentials --- a small enhancement of the in-plane ($V_\pi$) and a partial suppression of the out-of-plane ($V_\sigma$) components (full parameter values are reported in Table~\ref{tab:quanty}) --- consistent with the reduced interlayer bonding in the 2D limit. Crucially, on SiO$_2$ the excitations remain Raman-like, retaining well-defined energy losses --- the fingerprint of a still-localized, correlated-insulating ground state.

On the Au substrate (S2), the RIXS response changes dramatically, as shown in Fig.~\ref{fig:fig6}(c). The $d$--$d$ excitations are strongly broadened and washed out, and the multiplet structure of both the bulk and the SiO$_2$ bilayer is lost: a broad, featureless continuum with no resolvable gap replaces the discrete peaks, and spectral weight is massively transferred to the charge-transfer region. This is consistent with a fluorescence-like response of a band-like system, arising from a virtually ungapped electron--hole continuum, as established for correlated metals~\cite{bertinshaw2021ca3ru2o7,minola2017crossover}, and matches the metallic XAS lineshape of Fig.~\ref{fig:fig3}(c). 

The collapse of the measured localized electronic excitations marks the breakdown of the atomic multiplet description, and is further evidence of the interface-driven crossover toward a more itinerant regime --- the same crossover corroborated, on the same flakes, by the metallic XAS shape, the reduced $m_l/m_s$ ratio, and the strongly enhanced, field-linear XMCD response.

\section{Discussion}
Uncompensated magnetic effects have been reported at FePS$_3$/metal interfaces, including an anomalous Hall signal in a Pt layer~\cite{ma2025van} and interfacial Fe moments coupling antiferromagnetic FePS$_3$ to a ferromagnetic Co film~\cite{dhanarajgopal2021interfacial}. More broadly, charge transfer and electron doping are known to reshape its magnetic ground state, from substrate-induced modifications~\cite{lee2023giant} to intercalation-driven ferrimagnetic and hard-ferromagnetic order~\cite{mi2025electrical,ou2025hard}. The present study resolves the interfacial moment directly and element-specifically on the Fe sites, tracks it layer by layer, and traces it to its microscopic mechanism: a charge-transfer-driven reconstruction of the Fe~$3d$ states.

The comparison across our three samples isolates the role of the substrate unambiguously. Measured under identical conditions, the bulk shows only a weak field-induced uncompensation expected from a collinear antiferromagnet; the bilayer on inert SiO$_2$, a moderate enhancement attributable to the intrinsic response of the reduced-dimensional spin system~\cite{feringa2022observation,chaturvedi2025ab}; and the bilayer on gold, twice that value, rising to three times for the monolayer. With thickness, field, geometry and temperature held fixed in the bilayer comparison, this enhancement can be assigned to the metallic interface alone.

The atomic multiplet calculations describe quantitatively the two localized cases of this comparison: reproducing XAS, XMCD and RIXS with a single parameter set, they give access to the energy parameters (i.e., crystal field, trigonal distortion, charge-transfer energy and hybridization potentials) that define the Fe$^{2+}$ $3d$ manifold and its spin and orbital state, in the bulk as in the more covalent SiO$_2$ bilayer.

The origin of the gold-induced magnetic enhancement is instead electronic rather than chemical: with no significant shift of the Fe~$L_3$ edge, the loss of multiplet structure in both XAS and RIXS signals an increased hybridization of the Fe~$3d$ states with the Au bands, which transfers spectral weight toward a more itinerant-metallic regime and, as shown by the sum-rule analysis, partially quenches the orbital moment. That the Fe~$3d$ manifold of FePS$_3$ can become itinerant is well established under other tuning knobs. Pressure drives this Mott insulator through an insulator--metal transition~\cite{haines2018pressure}, orbital-selective in character~\cite{kim2022orbital}, at a critical pressure that collapses from $\sim$10.8~GPa in the bulk to $\sim$1.5~GPa in trilayers~\cite{mallick2024pressure}. Electron doping acts on the same manifold: the donated charge enters the transition-metal $d$ states in FePS$_3$, CoPS$_3$ and NiPS$_3$, as shown by photoemission spectroscopy and DFT+$U$~\cite{Nitschke2026}. The probed interfacial state can be viewed as the same crossover, reached through charge transfer from the contacting metal, the interfacial analogue of the Mottness collapse observed in layered correlated systems~\cite{qiao2017mottness}. RIXS traces it directly from the Raman-to-fluorescence collapse of the electronic excitations.

The giant single-ion anisotropy of FePS$_3$ stems from its large unquenched orbital moment~\cite{lee2023giant}, which relies on strong on-site correlations~\cite{ovesen2026giant}. The partial quenching of the orbital moment upon hybridization with gold implies a reduced single-ion anisotropy. In this partially itinerant regime, a Stoner-like enhancement of the interfacial $3d$ susceptibility may also contribute. Itinerant magnetism has several precedents in van der Waals systems --- Fe$_3$GeTe$_2$ being the archetypal itinerant ferromagnet down to the monolayer~\cite{fei2018two} --- and Stoner exchange, competing with superexchange, has been invoked for the ferrimagnetic net moment of NiPS$_3$ electron-doped by molecular intercalation~\cite{mi2022variation}. Consistently, the element-specific magnetization of the gold-supported flakes --- linear, non-hysteretic, and unsaturated up to 9~T --- points toward an enhanced field-induced susceptibility rather than spontaneous ferromagnetic order. The layer dependence sets the depth scale of this picture: the induced susceptibility is concentrated at the contact layer but not confined to it, decaying as interlayer antiferromagnetic exchange progressively outweighs the interfacial hybridization.

A final remark concerns the nature of the interface itself. Although often regarded as a chemically and physically inert support, gold does not leave the magnetism unchanged: it preserves the Fe valence, yet the same Au--S affinity that makes large-area mechanical exfoliation possible also provides an electronic pathway capable of reconstructing the magnetic and electronic ground state of the layers in contact. Since the affected depth is then comparable to the total thickness of a genuinely few-layer device, this may deserve consideration when interpreting magnetotransport across metal/MPS$_3$ interfaces~\cite{ma2025van,feringa2022observation,MnPS3}.


\section{Conclusions}
We have shown how a thin gold interface transforms the nearly compensated antiferromagnetic ground state of monolayer FePS$_3$ into one carrying a sizeable interfacial moment, with a magnetic response at 9~T and 5~K more than twice larger at matched thickness. Combining layer-resolved XMCD and RIXS on the same flakes with atomic multiplet calculations, we traced this emergent magnetism to interfacial charge transfer and Fe~$3d$--Au hybridization, which push the contact layer from the localized, correlated-insulating regime of the bulk toward a more metallic one. The effect is strongest at the Au-contacted layer and decays over the first three layers, making layer number an atomically precise handle on this interfacial magnetic moment. More generally, these results highlight substrate engineering as a deterministic route to control magnetization in van der Waals antiferromagnets.
Methodologically, combining Au-assisted exfoliation with on- and off-resonance X-ray absorption mapping makes element-specific XMCD and RIXS more accessible in the genuine few-layer limit, where flake size and selection pose practical difficulties. The approach transfers directly to the wider MPS$_3$ family and to other 2D magnets, opening the study of increasingly complex van der Waals heterostructures and interfaces.
\section*{Methods}
\begin{table*}[t]
\centering
\caption{Parameters used in the atomic multiplet calculations for the Fe~$L_{3,2}$-edge XAS, XMCD, and RIXS simulations of bulk and bilayer FePS$_3$. Values in bold differ between the two cases. All values are expressed in eV, except the dimensionless Slater-integral scaling factor $\beta$.}
\label{tab:quanty}
\small
\renewcommand{\arraystretch}{1.25}
\begin{tabular*}{\textwidth}{@{\extracolsep{\fill}}lcccccccccccccc}
\hline\hline
& \multicolumn{5}{c}{$\mathcal{H}_{\mathrm{atomic}}$}
& \multicolumn{2}{c}{$\mathcal{H}_{\mathrm{CF}}$}
& \multicolumn{5}{c}{$\mathcal{H}_{\mathrm{LMCT}}$}
& $\mathcal{H}_{\mathrm{lig}}$
& $\mathcal{H}_{\mathrm{mag}}$ \\
\cline{2-6} \cline{7-8} \cline{9-13} \cline{14-14} \cline{15-15}
& $U_{dd}$ & $U_{pd}$ & $\zeta_{3d}^{\mathrm{init}}$ & $\zeta_{3d}^{\mathrm{fin}}$ & $\beta$
& $10Dq$ & $\tau$
& $\Delta$ & $\Delta_{\mathrm{interm}}$ & $V_\sigma$ & $V_\pi$ & $V_{a_{1g}}$
& $10Dq_{\mathrm{lig}}$
& $g_s\mu_B\mu_0 H_E$ \\
\hline
Bulk
& \textbf{4.5} & \textbf{6.4} & 0.058 & 0.074 & \textbf{0.78}
& \textbf{1.00} & 0.15
& \textbf{1.6} & \textbf{1.1} & \textbf{1.2} & \textbf{1.0} & 1.1
& 0.40
& 0.035 \\
Bilayer
& \textbf{4.0} & \textbf{5.7} & 0.058 & 0.074 & \textbf{0.75}
& \textbf{0.95} & 0.15
& \textbf{0.8} & \textbf{0.5} & \textbf{0.8} & \textbf{1.1} & 1.1
& 0.40
& 0.035 \\
\hline\hline
\end{tabular*}
\end{table*}

\subsection*{Exfoliation and sample preparation}
Few-layer FePS$_3$ flakes were obtained by mechanical exfoliation from bulk single crystals (HQ Graphene~\cite{hqgraphene}) onto two types of substrate: bare SiO$_2$/Si (oxide thickness 300~nm) and Au(2--4~nm)/Ti(2~nm)/SiO$_2$/Si. The Ti adhesion layer and the Au film were deposited by electron-beam evaporation (Plassys MEB550S) at a rate of 0.1~\AA/s and a base pressure below $10^{-6}$~mbar. Substrates were activated by O$_2$ plasma for one minute immediately before deposition, and the freshly evaporated gold was used within 5~minutes, since the efficiency of the technique is highly sensitive to contamination of the gold surface~\cite{desai2016gold,huang2020universal,velicky2018mechanism}. Samples S1 and S2 were exfoliated inside an argon-filled glovebox, whereas sample S3 was exfoliated in a cleanroom under ambient conditions and transferred to the glovebox immediately afterwards. The position of each atomically thin region of interest (ROI) was recorded using a coordinate system referenced to a corner of the substrate defined as the origin $(0,0)$; this allowed the X-ray beam to be positioned close to each ROI at the beamline, where the flakes were then precisely located by X-ray absorption mapping. Further details on the flake-location procedure are given in the Supplementary Information.

\subsection*{Raman scattering spectroscopy, AFM and magnetometry}
Raman scattering spectra were acquired with a Horiba T64000 apparatus in single-monochromator configuration, using a green excitation laser (532~nm) focused to a spot of $\sim$2~$\mu$m in diameter, at a power below 1~mW to avoid laser-induced heating and degradation. In this configuration, the spectral precision is about 1~cm$^{-1}$, which sets the uncertainty of the Raman peak positions reported here. For air-stability tests, samples were transferred and initially measured in a sealed quartz cuvette to prevent uncontrolled air exposure; stability was then assessed by exposing the sample to air and repeating the acquisition on the same spot at successive time intervals. Atomic force microscopy was carried out with a Bruker Dimension Icon in tapping mode, used to determine the thickness of the air-exfoliated flakes and to confirm the thickness of the synchrotron-measured flakes after the beamtime. The complete AFM and Raman characterization, including the air-stability study, is reported in the Supplementary Information. Magnetization measurements were performed on bulk FePS$_3$ crystals using a VSM-SQUID magnetometer. Magnetization was measured as a function of temperature from 2 to 300~K under an out-of-plane field of 0.5~T, and as a function of field while sweeping between $-6$ and $+6$~T at fixed temperatures.

\subsection*{Synchrotron X-ray measurements}
XMCD measurements were carried out at the European Synchrotron Radiation Facility (ESRF) on beamline ID32~\cite{ID32,brookes2018beamline} at the Fe~$L_{3,2}$ absorption edges, using circularly polarized soft X-rays at the High-Field Magnet (HFM) endstation (base pressure around $5\times10^{-10}$~mbar). The incident beam exhibited a degree of circular polarization close to 100\% and a footprint of approximately FWHM $10\times20~\mu$m$^2$ on the sample.

Flakes were located in situ by X-ray absorption mapping at on- and off-resonance energies (Fe~$L_3$ resonance and 700~eV); the correspondence between the resulting maps and the optical micrographs allowed the beam to be positioned on regions of known thickness. For each thickness, several regions were measured on different flakes and on different samples, to verify the reproducibility of the XAS and XMCD lineshapes.

X-ray absorption was recorded in total electron yield (TEY) mode and normalized to the incident flux, measured with a diamond thin film drain current~\cite{Kummer2013}. The XMCD spectra shown were obtained as the difference between X-ray absorption spectra recorded with left- and right-circularly polarized light, and are reported as a percentage of the average XAS edge jump. To suppress field-independent artifacts, the XMCD signal was antisymmetrized with respect to magnetic-field reversal, $\mathrm{XMCD}(H)=\tfrac{1}{2}\left[\mathrm{XMCD}(+H)-\mathrm{XMCD}(-H)\right]$, and the uncertainty was estimated from the residual field-independent contribution, $\mathrm{error}(H)=\tfrac{1}{2}\left[\mathrm{XMCD}(+H)+\mathrm{XMCD}(-H)\right]$. All measurements were performed at $5\pm0.5$~K under magnetic fields from 0 to 9~T applied parallel to the incident beam. In the chosen normal-incidence geometry --- the best compatible with precise positioning on the flakes --- the beam and field are directed along the $c$ axis, i.e.\ the van der Waals stacking direction. Beamline flux and polarization were routinely monitored to ensure reproducibility across magnetic-field cycles, and energy shifts were checked against a reference Fe$_2$O$_3$ powder absorption spectrum.

Orbital-to-spin moment ratios $m_l/m_s$ were obtained from the magneto-optical sum rules~\cite{thole1992x,carra1993x} applied to the integrated XMCD spectra, neglecting the magnetic dipole term $\langle T_z\rangle$ (see Supplementary Information); being independent of the number of $3d$ holes and of the XAS normalization, the ratio is compared across samples measured and analysed under identical conditions.

RIXS measurements were performed at the same ID32 beamline using the dedicated RIXS spectrometer integrated into the HFM endstation, enabling acquisition under the same low-temperature and sample normal incidence geometry conditions as the XMCD measurements, in zero applied field. The spectrometer operates at a scattering angle of $90^{\circ}$, corresponding to a fixed transferred momentum $|\mathbf{q}|\approx0.51$~\AA$^{-1}$ at the Fe~$L_3$ resonance ($\sim$707~eV), and emitted photons were detected with a CMOS detector in single-photon-counting (SPC) mode. The energy maps of Fig.~\ref{fig:fig5}(a) were acquired with the average of horizontal- and vertical-linearly polarized light, while single spectra were acquired with horizontal incident polarization, which partially suppresses the elastic and diffuse scattering and enhances the contrast of the measured electronic excitations. The total energy resolution was around 220~meV at the Fe~$L_{3}$ edge, sufficient to resolve intra-atomic $d$--$d$ transitions and charge-transfer excitations, though not lower-energy excitations (i.e.\ phonons, magnons, and the spin--orbit exciton).

\subsection*{Atomic multiplet calculations}
X-ray absorption, XMCD, and RIXS spectra at the Fe~$L_{3,2}$ edges were simulated within the Anderson impurity model as implemented in the many-body package \textsc{Quanty}~\cite{haverkort,quanty}. The calculations are based on a fully parameterized Hamiltonian for a single Fe$^{2+}$ cluster, in which ligand-field effects are explicitly included through charge-transfer and hybridization terms up to the $d^{7}\underline{L}$ configuration (single ligand hole). The local symmetry of the cluster is octahedral ($O_h$) with a trigonal distortion ($D_{3d}$). The total Hamiltonian reads
\begin{equation}
\mathcal{H}_\mathrm{tot} = \mathcal{H}_\mathrm{atomic} + \mathcal{H}_\mathrm{CF} + \mathcal{H}_\mathrm{LMCT} + \mathcal{H}_\mathrm{lig} + \mathcal{H}_\mathrm{mag},
\label{eq:hamiltonian}
\end{equation}
where $\mathcal{H}_\mathrm{atomic}$ contains the intra-atomic Coulomb repulsions between $3d$ electrons ($U_{dd}$) and between a $3d$ electron and a $2p$ core hole ($U_{pd}$), together with the $3d$ spin--orbit coupling in the initial and final states ($\zeta_{3d}^{\mathrm{init}}$, $\zeta_{3d}^{\mathrm{fin}}$); the atomic Slater integrals are uniformly reduced by a factor $\beta$ to account for intra-atomic screening. $\mathcal{H}_\mathrm{CF}$ describes the octahedral and trigonal crystal field ($10Dq$, $\tau$); $\mathcal{H}_\mathrm{LMCT}$ the metal--ligand hybridization potentials ($V_\sigma$, $V_\pi$, $V_{a_{1g}}$) and the charge-transfer energy ($\Delta$) in the initial and intermediate states; $\mathcal{H}_\mathrm{lig}$ the ligand crystal field ($10Dq_\mathrm{lig}$); and $\mathcal{H}_\mathrm{mag}$ the magnetic term. The latter accounts for both the applied external field (Zeeman term, $\mu_B\,\mu_0 H_\mathrm{ext}\cdot(\mathbf{L}+2\mathbf{S})$), set to the experimental value of each measurement, and an effective internal molecular mean field ($g_s\mu_B\,\mu_0 H_E\cdot\mathbf{S}$) describing the exchange interaction.
To account for the two antiferromagnetic sublattices, the spectra are computed for two clusters under opposite molecular exchange fields, $\pm H_E$, along the $c$ axis. Equal weights describe the compensated antiferromagnet, while an increased weight of the field-aligned cluster models a field-induced sublattice uncompensation as discussed in the main text. Calculated spectra were broadened by a Lorentzian of 0.3~eV (FWHM) at the $L_3$ edge and 0.5~eV at the $L_2$ edge to account for the core-hole lifetime, and by a Gaussian of 0.1~eV reproducing the experimental resolution. Starting from parameter values reported in the literature for FePS$_3$~\cite{wei2025spin}, we systematically varied the model parameters to best reproduce the measured spectra. The optimized parameter sets for the bulk and for the atomically thin flake on SiO$_2$ are reported in Table~\ref{tab:quanty}. Further details are given in the Supplementary Information.

\section*{Data Availability}

The raw XAS/XMCD/RIXS data supporting the findings of this study are available at \url{https://doi.org/10.15151/ESRF-ES-2411085802} (an embargo may apply).
The raw Raman, AFM, and magnetometry data, as well as the \textsc{Quanty} scripts used for the calculations, are available from the corresponding author upon reasonable request.

\section*{Acknowledgements}

The authors acknowledge the European Synchrotron Radiation Facility (ESRF) for the provision of beamtime at the ID32 beamline under proposal number IHHC4284, and are grateful to Dr.\ Flora Yaokhu Harris and Ms.\ Céline Allais, beamline staff at ID32, for their technical support during the beamtime. The authors thank Simon Le-Denmat for assistance with the AFM measurements and Didier Dufeu for support with the magnetometry measurements at the Institut Néel--CNRS (Grenoble). The authors also thank HQ Graphene for providing the bulk crystals from which the exfoliated samples were prepared, and the cleanroom staff at the Institut Néel for their technical support during sample fabrication. This work is also supported by France 2030 government investment plan managed by the French National Research Agency under grant reference PEPR SPIN - SPINMAT ANR-22-EXSP-0007. 

\section*{Author Contributions}

G.B.\ led the experimental work, actively took part in all stages of the project from sample preparation to measurements, performed the data analysis and simulations, and wrote the original draft of the manuscript. A.P.\ fabricated the exfoliated samples, carried out the AFM and Raman characterization, participated in the synchrotron beamtime, performed data analysis, and assisted with manuscript editing. M.d.R.C.\ and R.N.\ contributed to the preparation and execution of the synchrotron beamtime, participated in both the XMCD and RIXS measurements, and helped interpret the results. A.A.\ supported sample preparation and characterization and contributed to the development of the sample-preparation strategy. R.S.\ participated in the synchrotron beamtime and contributed to the atomic multiplet calculations. T.R.\ supported sample preparation and characterization and performed the magnetometry measurements. V.R., N.Be.\ and L.M.\ provided support for setting up the Raman spectroscopy instrumentation, assisted during the Raman measurements, and facilitated the collaboration between Institut N\'{e}el and the ESRF. K.K.\ provided support during the beamtime and helped develop and interpret the results and atomic multiplet calculations. J.C.\ secured funding and access through Institut N\'{e}el and actively participated in the scientific interpretation of the results. N.B.B.\ conceived and supervised the project at all stages, secured funding, participated in the beamtime, and actively contributed to the scientific interpretation of the results. All authors contributed to the scientific discussion, revised the manuscript, and approved the final version.

\section*{Competing Interests}

The authors declare no competing interests.

\bibliographystyle{naturemag}
\bibliography{reference}


\newpage
\onecolumn

\end{document}